\input{aipcheck}

\documentclass[
    ,final            
  ]
  {aipproc}

\layoutstyle{6x9}

\begin{document}

\title{Chiral symmetry, scalar field and 
confinement : from nucleon 
structure to nuclear matter
}


\author{Guy Chanfray}{
  address={IPN Lyon, IN2P3/CNRS, Universit\'e Lyon1, France}
}

\author{Magda Ericson}{
  address={Theory Division, CERN, Geneva, Switzerland}
}


\begin{abstract}
We discuss the relevance of the scalar modes appearing in chiral theories with spontaneous symmetry breaking such as the NJL model for nuclear matter studies. We show that it depends on the relative role  of  chiral symmetry breaking and confinement in the nucleon mass origin. It is only in the case of a mixed origin  that nuclear matter can be stable and reach saturation.
We describe models of nucleon structure where this balance is achieved. We show how chiral constarints and confinement modify the QCD sum rules for the mass evolution in nuclear matter.    
\end{abstract}

\maketitle


\subsection{Introduction }
 One  crucial question of present day nuclear physics is the interplay between  the nuclear many-body problem and the nucleon structure where both chiral symmetry breaking and confinement are involved. In this talk we will develop  a viewpoint where the nuclear attraction is associated with the in-medium fluctuation of a (chiral invariant) background scalar field which is itself at the origin of the constituent quark mass. Nuclear stability is ensured with the incorporation  of the nucleon  response to this  scalar field. This  response  depends on the quark confinement mechanism inside the nucleon. As we will see this question can be  rephrased in terms of the relative weight of spontaneous chiral symmetry breaking and confinement in the origin of the nucleon mass.

\subsection{Chiral nuclear matter description including the effect of confinement}
As a starting point let us consider  the relativistic mean-field approaches initiated by Walecka \cite{SW86} where the nucleons move in an attractive scalar  and a repulsive vector background fields. It provides an economical saturation mechanism and a well known success is the correct magnitude of the spin-orbit potential where  the large vector and scalar fields contribute  in  an additive way. Now the question of the very nature of these background fields  and their  relationship with the QCD condensates has to be elucidated. To address this question we formulate an effective theory     parametrized in terms of the fields associated with the fluctuations of the chiral condensate in a matrix form (${\cal M}=\sigma + i\vec\tau\cdot\vec\pi$) 
by going from cartesian  to polar coordinates,  {\it i.e.}, going from a linear to a non linear representation~:
${\cal M}=\sigma\, + \,i\vec\tau\cdot\vec\pi=S\,U=(f_\pi\,+\,s)\,exp\left({i\vec\tau\cdot\vec\Phi_\pi/ f_\pi}\right)$.  In ref. \cite{CEG02} we made the physical assumption  to identify the chiral invariant scalar field, $s=S - f_\pi$, associated with {\it radial} (in order to respect chiral constraints) fluctuations   of the condensate, with the background attractive scalar field. 
The normalization is such that the symmetry breaking piece of the Hamiltonian is given by $H_{\chi SB}=-\int d^3 r\,F_\pi M^{2}_{\pi}\,\sigma$.
It follows  that the evolution of the condensate is given by~:
\begin{equation}
 \frac{\left\langle \bar{q}q\right\rangle}{\left\langle \bar{q}q \right\rangle_{vac}} \simeq
1\,+\,\frac{\bar{s}}{F_\pi}\,-\,\frac{\left\langle \Phi^2\right\rangle}{2\,F^{2}_\pi}\simeq 1\,-\,\frac{\sigma_N\,\rho}{F^2_\pi\,M^{2}_{\pi}}\label{COND}
\end{equation} 
The first equality is valid to leading order in the pion field and the second equality holds for a dilute system.
The quantity  $\bar s=\langle s \rangle $ is the expectation value of the nuclear scalar field. In this picture  the
nuclear medium can be seen as a shifted vacuum characterized by a chiral order parameter $\bar{S}= F_\pi + \bar{s}$ and the nucleon mass will depend in some way on the scalar field $M^*_N=M^*_N(S)$. The dynamics of the scalar field is encoded in the chiral effective potential, $W(s)={(M^{2}_{\sigma})_{eff}}\,s^2/2 +\,....$, associated with chiral symmetry breaking. Notice that the effective sigma mass has no reason to coincide with the mass of the $f_0(600)$ \cite{CSWSX95,CWS01}. In the low density limit the expectation value of the scalar field is 
$\bar{s}=-g_\sigma\rho/{(M^{2}_{\sigma})_{eff}}$ where $g_\sigma=\left(\partial M_N/\partial s\right)_{s=0}$ is the scalar coupling of the nucleon. The pion-nucleon sigma sigma commutator  receives a contribution from the scalar field and one from the pion cloud, 
$\sigma_N= \sigma_{N\sigma}^{(nopion)}+\sigma_N^{(pion\,cloud)}$~:
\begin{equation}
\sigma^{(no pion)}_{N \sigma}=F_\pi g_\sigma\,\frac{M^{2}_{\pi}}{(M^{2}_{\sigma})_{eff}},\qquad
\sigma_N^{(pion\,cloud)}= \int d^3 r\,\frac{1}{2} M^{2}_{\pi}\,\left\langle N\right|\Phi^2\left|N\right\rangle .
\end{equation}
The quark condensate evolution can be compared with the one of the nucleon mass which, for a dilute system, is~:
\begin{equation}
\frac{M^*_N}{M_N}\simeq   1\,+\,\frac{\bar{s}}{M_N}\simeq  1\,-\,\frac{g_\sigma F_\pi}{M_N}\,\frac{\sigma_{N \sigma}^{(nopion)}\rho}{F^2_\pi\,M^{2}_{\pi}}\simeq
1\,-\,\frac{g_\sigma}{10}\,\frac{\sigma_{N\sigma}^{(nopion)}\rho}{F^2_\pi \,M^{2}_{\pi}}\label{MASS} .
\end{equation}
We see that this result deviates from the Ioffe sum rule \cite{I83} which predicts that the quark condensate and the nucleon mass 
have the same evolution with density controlled by the pion nucleon sigma term \cite{CFG91,FKVW06}. Firstly the pion cloud piece of the sigma commutator, $\sigma_N^{(pion\,cloud)}$,  contributes to the condensate evolution. It does not contribute to the mass evolution, otherwise chiral constraints would be violated (such as the presence of a term in $m_{\pi}$ in the NN potential, forbidden \cite{BM92} by chiral symmetry). In fact its influence on the mass vanishes in the chiral limit and hence it is a small effect which we ignore.  Only the scalar piece,  $\sigma_{N \sigma}^{nopion}$, should then enter the mass evolution. There is however another deviation from the Ioffe sum rule which is the presence of the weighting factor, $g_\sigma F_\pi/M_N$, where the scalar couping constant,  $g_\sigma$, depends on the nucleon structure.  Only in case where the nucleon mass entirely originates from chiral symmetry, such as in the linear sigma model, or naive additive NJL model ($ g_\sigma =M_N/F_\pi$), the Ioffe sum rule 
is recovered but corrected from pionic effect. The pion-nucleon sigma term entering these evolutions is an important piece of experimental information. It is obtained from the Feynman-Hellman theorem~:
$\sigma_N=m \,(\partial M_N/\partial m) \simeq 50\,MeV$. According to  previous works \cite{BM92,LTY04,CE07}, we expect for the pion cloud contribution $\sigma_N^{(pion\, cloud)}\simeq 20\, MeV$. The numerical value of the non pionic piece    has to be $\sigma^{(no pion)}_{N\sigma} = \sigma_N  -\sigma_N^{( pion\,cloud)} \simeq 50 - 20\simeq 30\, MeV$. We will come back to this important constraint later on.

The Hartree energy density of nuclear matter (including omega exchange) writes in terms of the order parameter $\bar s$ as~:
$\,\quad E_0/ V=\varepsilon_0=\int\,(4\,d^3 p /(2\pi)^3) \,\Theta(p_F - p)\,E^*_p(\bar s)\,+\,W(\bar s)\,+\,g^2_\omega/2m_\omega^2\,\rho\,$
where $E^*_p(\bar s)=\sqrt{p^2\,+\,M^{*2}_N(\bar s)}$ is the energy of a nucleon with momentum ${\bf p}$.  Here two serious problems appear when the nucleon mass has a pure symmetry breaking origin~: $M^{*}_{N}(\bar s)=M_N + g_{\sigma}\bar s$. The first one is  the fact that the chiral effective potential, $W(s)$, contains an attractive tadpole diagram which  generates an attractive three-body force destroying  matter stability \cite{BT01}. The second one is related to the nucleon substructure. According to the lattice data analysis of  Thomas et al \cite{LTY04}, the nucleon mass can be expanded according to $M_N(m^{2}_{\pi}) = a_{0}\,+\,a_{2}\,m^{2}_{\pi}\, +\,a_{4}\,m^{4}_{\pi}\,+\,\Sigma_{\pi}(m_{\pi}, \Lambda)+...$, where the pionic self-energy is explicitly separated out. While the $a_2$ parameter is related to the non pionic 
piece of the $\pi N$ sigma term,  $a_4$ is related to the nucleon QCD scalar susceptibility. The important point is that   $(a_4)_{latt} \simeq- 0.5\, \mathrm{GeV}^{-3}$ \cite{LTY04} is essentially compatible with zero in the sense that it is  much smaller than in our chiral effective model, $\left(a_4\right)_{Chiral}=-3F_{\pi} g_\sigma/2\,(M^{4}_{\sigma})_{eff}\simeq -3.5\,GeV^{-3}$ , where the nucleon  is seen  as a juxtaposition of three constituent quarks getting their mass  from the chiral condensate \cite{CE05}. The common origin of these two failures can be attributed to the absence  of confinement \cite{MC08}. In reality the composite nucleon  responds to the nuclear environment, {\it i.e.}, by readjusting its confined quark structure \cite{G88}. The resulting  polarization of the nucleon is accounted for by the phenomenological introduction of the {\it positive} scalar nucleon response, $\kappa_{NS}$, in  the nucleon mass evolution ~: $M_N(s)=M_N\,+\,g_\sigma\,s\,+\,\frac{1}{2}\,\kappa_{NS}\,s^2\,+\,....$.
This constitutes the only change in the expression of the energy density  but this has numerous consequences. In particular
the $a_4$ parameter is modified~: $a_4=\left(a_4\right)_{Chiral}\left(1\,-\,\frac{2}{3} C\right)$. The value of  $C\equiv(F^{2}_{\pi}/2\,M_N)\kappa_{NS}$ which reproduces the lattice data is $C\simeq 1.25$ implying a strong cancellation effect in $a_4$
\cite{CE05, CE07}. Moreover  the scalar response of the nucleon induces an new piece in the lagrangian ${\cal L}_{s^2 NN}=-\,\kappa_{NS}\,s^2\,\bar N N/2$ which  generates a repulsive three-body force able to restore saturation \cite{EC07}.

The restoration of saturation properties has been confirmed at the Hartree level \cite{CE05} with a value of the dimensionless scalar response parameter, $C$, close to the value estimated from the lattice data, taking   $g_\sigma=M_N/F_\pi$ and $g_\omega$ adjusted near  the VDM value. The next step has been to include pion loops on top of the Hartree mean-field calculation \cite{CE07,MC09}. One possibility is to use in-medium chiral perturbation theory but we prefer to use a standard many-body (RPA) approach which includes the effect of short-range correlations ($g'$ parameters fixed by spin-isospin phenomenology) and  $\Delta-h$ excitations.   The calculation which also incorporates rho exchange has no real free parameters apart for a fine tuning of $C$ (around the lattice estimate) and of  $g_\omega$  (around the VDM value). The inclusion of spin-isospin  loops improves the quality of the result,  in particular  they bring down the compressibility close to the accepted value. We stress   the relatively modest value of the correlation energy ($\simeq -10\,MeV$), much smaller than what is obtained from iterated pion exchange (planar diagramm) in  in-medium chiral perturbation theory. This effect is mainly due to the strong screening of pion exchange by short-range correlations. For completeness we mention that

we have also performed  a full relativistic Hartree-Fock calculation with the notable inclusion of the rho meson exchange which allows to also reproduce asymmetry properties of nuclear matter. Details can be found in ref. \cite{MC08}.

\subsection{Status and relevance of the nuclear scalar field}
Although the concept of a  scalar  field  has been widely used for nuclear matter studies \cite{SW86}  its precise origin or meaning  is still a controversial subject. The problem is that there is no sharp scalar resonance which would lead to a simple scalar particle exchange. In our approach instead we stress the chiral aspect of the problem. As soon as we start from a model which gives a correct description of chiral symmetry breaking in the QCD vacuum such as the Nambu-Jona-Lasinio model (NJL), the emergence of a scalar field linked to the quark condensate cannot be avoided. 
 For this purpose we introduce the NJL model in  the light quark sector limited to describe the main mesons~:  the pion, the sigma, the rho, the $a_1$ and the omega mesons. The  lagrangian is~:
$
{\cal L}= \bar{\psi}\left(i\,\gamma^{\mu}\partial_\mu\,-\,m\right)\,\psi\,+\,(G_1/2)\,\left[\left(\bar{\psi}\psi\right)^2\,+\
\left(\bar{\psi}\,i\gamma_5\vec\tau\,\psi\right)^2\right]
 \,-\,(G_2/2)\,\left[\left(\bar{\psi}\,\gamma^\mu\vec\tau\,\psi\right)^2\,+\,
\left(\bar{\psi}\,\gamma^\mu\gamma_5\vec\tau\,\psi\right)^2\,+\,\left(\bar{\psi}\,\gamma^\mu\,\psi\right)^2\right]\,$.
It is generally accepted that this model gives an excellent description of vacuum chiral symmetry breaking, with the appearence of a constituent quark mass and a soft Goldstone pion mode. In addition in the so called delocalized version the inclusion of  a form factor, depending on a cutoff, $\Lambda$, at each quark leg of the interaction generates a momentum dependence of the quark mass in agreement with lattice calculation. However the approach  is notoriously not fully satisfactory due to the lack of confinement~: in particular unphysical decay channels of vector mesons in $q \bar q$ pairs may appear. In that respect  it has been shown that adding a confining interaction on top of the NJL model solves the problem of unphysical $\bar q q$ decay channels of mesons with masses larger than twice the constituent quark mass \cite{CSWSX95}. In addition the scalar meson arising around twice the constituent quark mass is pushed at  higher energy by confinement and the $\sigma(600)$ comes out as a broad $\pi\pi$ resonance. It has also been demonstrated that the effect of the confining interaction, although crucial for the on-shell properties of vector and scalar mesons, has only little influence for the low momentum physics  relevant for the nuclear many-body problem \cite{CWS01}. Hence, our attitude will be to derive an effective low momentum theory where the mass parameters for scalar et vector mesons will not be the on-shell mesons masses but simply mass parameters associated with the inverse of the corresponding correlators taken at zero momentum. Technically this can be done by  rewriting the NJL lagrangian in a semi-bozonized form
and integrating out quarks in the Dirac sea using a path integral formalism. The physical meaning is simply a projection of $q \bar q$ vacuum fluctuations onto meson degrees of freedom. Keeping only the relevant terms for nuclear physics purpose, the resulting low momentum effective lagrangian has the form~:
\begin{eqnarray}
{\cal L}&=&\frac{1}{2}\,\frac{I^l_{2S}(\bar{\cal S})}{I^l_{2S}(M_0)}\partial^{\mu} S_c\partial{\mu}S_c\,-\,W	({\cal S}=g_{0S} S)
+\frac{1}{4}\,F^2\,M_\pi^2\,\frac{\bar{\cal S}}{M_0}\,tr_f(U\, +\, U^\dagger\,-\,2)\,\nonumber\\
& &+\,
\frac{1}{2 F^2}\tilde I(\bar{\cal S})\,\bar{\cal S}^2\,\partial^{\mu}\vec{\Phi}\partial^{\mu}\vec{\Phi}
+ \frac{1}{2}M^{2}_{V}\,\left(\omega^{\mu}\omega_{\mu}\,+\,\vec{v}^{\mu}\cdot\vec{v}_{\mu}\right)\,+\,
\frac{1}{2}M^{2}_{A}\,\left(\vec{a}^{\mu}\cdot\vec{a}_{\mu}\right)
\nonumber\\
& &-\,\frac{1}{4}\,\left(\omega^{\mu\nu}\omega_{\mu\nu}\,+\,\vec{v}^{\mu\nu}\cdot \vec{v}_{\mu\nu}\,+\,\vec{a}^{\mu\nu}\cdot \vec{a}_{\mu\nu}\right).
\end{eqnarray}
${\cal S}$ is a chiral invariant scalar field whose vacuum expectation value coincides with the constituent quark mass $M_0$ and $S_c$ is the canonical one. We also introduce an effective scalar field  $(S)_{eff}=(F_\pi/M_0){\cal S}\equiv F_\pi + s$  normalized to $F_\pi$ in the vacuum which coincides with the chiral invariant $S$  field introduced in the second section and  $W$ is the corresponding chiral effective potential. The matrix $U$ is $U=exp(\Phi/F)$ where $\Phi$ is the orthoradial pion field and the parameter $F$ can be identified with the pion decay constant parameter $F_\pi$. The quantity
$\bar{\cal S}=(M_0/F_\pi)(F_\pi +\bar s)$ is the expectation value of the scalar field which is expected to decrease in the nuclear medium. The various $I$'s functions are standard NJL loop integrals which also enter the mass parameters $M_V$ and $M_A$  for vector and axial-vector mesons. 
These masses also depend on the vector coupling constant $G_2$. In particular the ratio $g^{2}_{V}/M^{2}_{V}$ ($g_V$ being the quark-vector coupling constant) is fixed to $G_2$. Finally the pion mass emerges as $M^{2}_{\pi}=m\,M_0/G_1\,F^2_\pi$.
We have a priori four parameters,  $G_1, G_2$, the cutoff  $\Lambda$ and the bare quark mass $m$.  We use~: $ \Lambda=1\, GeV, \quad m=3.5\, MeV, \quad G_1=7.8\, GeV^{-2}.$
We thus obtain for the vacuum quark mass at zero momentum~:
$M_0= 371\, MeV$ and $\left\langle \bar q q\right\rangle= -(286\, MeV)^3$. We actually constrain $G_2$ to be close to the VDM value
$(G_2)^{VDM}= g^{2}_{V}/M^{2}_{V}=(2.65/0.770)^2\,GeV^{-2}$. For the nuclear matter calculation we take $G_2=0.78\,(G_2)^{VDM} \Rightarrow F_\pi=93.6 MeV, \quad M_\pi=137.8\, MeV  $. With this set of parameters the low momentum mass parameters are~:
$ (M_\sigma)_{eff}=653\, MeV$,  $M_V=1256\, MeV$  and $M_A=1398\, MeV$. 

In the pure NJL picture, at finite baryonic density, the value of the constituent quark mass, which is the expectation value of the scalar field ${\cal S}$, is modified. It can be obtained self-consistently from a  gap equation modified by the presence of a Fermi sea. 
However in the real world baryonic matter is not made of independant constituent quarks but of clustered objects, the nucleons. These nucleons are embedded in the scalar background field, $\bar{\cal S}$, and the nuclear medium can be seen, as said before, as a shifted vacuum. The nucleon mass will depend in some way on the scalar background field and the energy density of  symmetric nuclear matter at the Hartree level reads~:
\begin{equation}
{E_0\over V}=\varepsilon_0=\int\,{4\,d^3 p\over (2\pi)^3} \,\Theta(p_F - p)\,\left(\sqrt{p^2+M^{2}_{N}(\bar{\cal S})}
\,-\,(M_N)_{vac}\right)
\,+\,W(\bar{\cal S})\,+\,9\,\frac{G_2}{2}\,\rho^2.\label{EOS}
\end{equation}
The expectation value for the scalar field is self-consistently obtained by minimization of the energy density,
\begin{equation}
	\frac{\partial\varepsilon_0}{\partial \bar{\cal S}}=0 \quad\Leftrightarrow\quad \frac{\bar{\cal S}- m}{G_1}=-2\,\left\langle \bar q q\right\rangle(\bar{\cal S})\,-\,\frac{\partial M_N}{\partial\bar{\cal S}}(\bar{\cal S})\,\rho^{N}_{s}(M_N(\bar{\cal S})),\label{MIN}
\end{equation}
which constitutes an in-medium modified gap equation with  
$\rho^{N}_{s}(M_N(\bar{\cal S}))$ being the nucleonic scalar density.
The scalar coupling constant of the nucleon to the effective scalar field (which is normalized to $F_\pi$ in the vacuum) is~:
$$(g_S)_{eff}(\bar{\cal S})=\frac{M_0}{F_\pi}\left(\frac{\partial M_N}{\partial \bar{\cal S}}\right)$$
which depends crucially on nucleon structure. For instance if the nucleon mass fully originates from confinement (bag  models), $\partial M_N/\partial \bar{\cal S} =0$, the scalar field just decouples from the nucleon, ($(g_S)_{eff}=0$). In this case  there is no shift of the vacuum and the scalar field is thus an irrelevant concept for nuclear matter studies. On the other extreme if the nucleon mass fully originates from chiral symmetry breaking (naive additive NJL, chiral soliton), then the nucleon mass in the medium is affected by the scalar field associated with the dropping of the chiral condensate. However in this case the attractive tadpole destroys stability. Only in the case where the nucleon mass has a mixed origin, the scalar background field can contribute to the nuclear attraction without destroying the stability and saturation properties. In that case 
 by rearranging its  quark structure linked to the confinement mechanism, the nucleon  reacts against the scalar field generating  effectively repulsive three-body forces. The origin of this repulsion lies in the decrease of the scalar coupling constant of the nucleon. In short a  possibly important   part of the saturation mechanism is associated with the progressive decoupling of the nucleon from the scalar field associated with the 
dropping of the chiral condensate. In the next section we will introduce nucleon models capable of achieving
the balance between large enough attraction and sufficient reaction. Of course, one falls here in the modelling
uncertainties. However there is  that a  stringent constraint  for the numerical value of the
scalar nucleon coupling constant which is model dependent, from the value of the free nucleon sigma commutator.

\noindent
\begin{figure}
  \begin{tabular}{cc}
  \begin{minipage}{.50\linewidth}
    \includegraphics[scale=0.3]{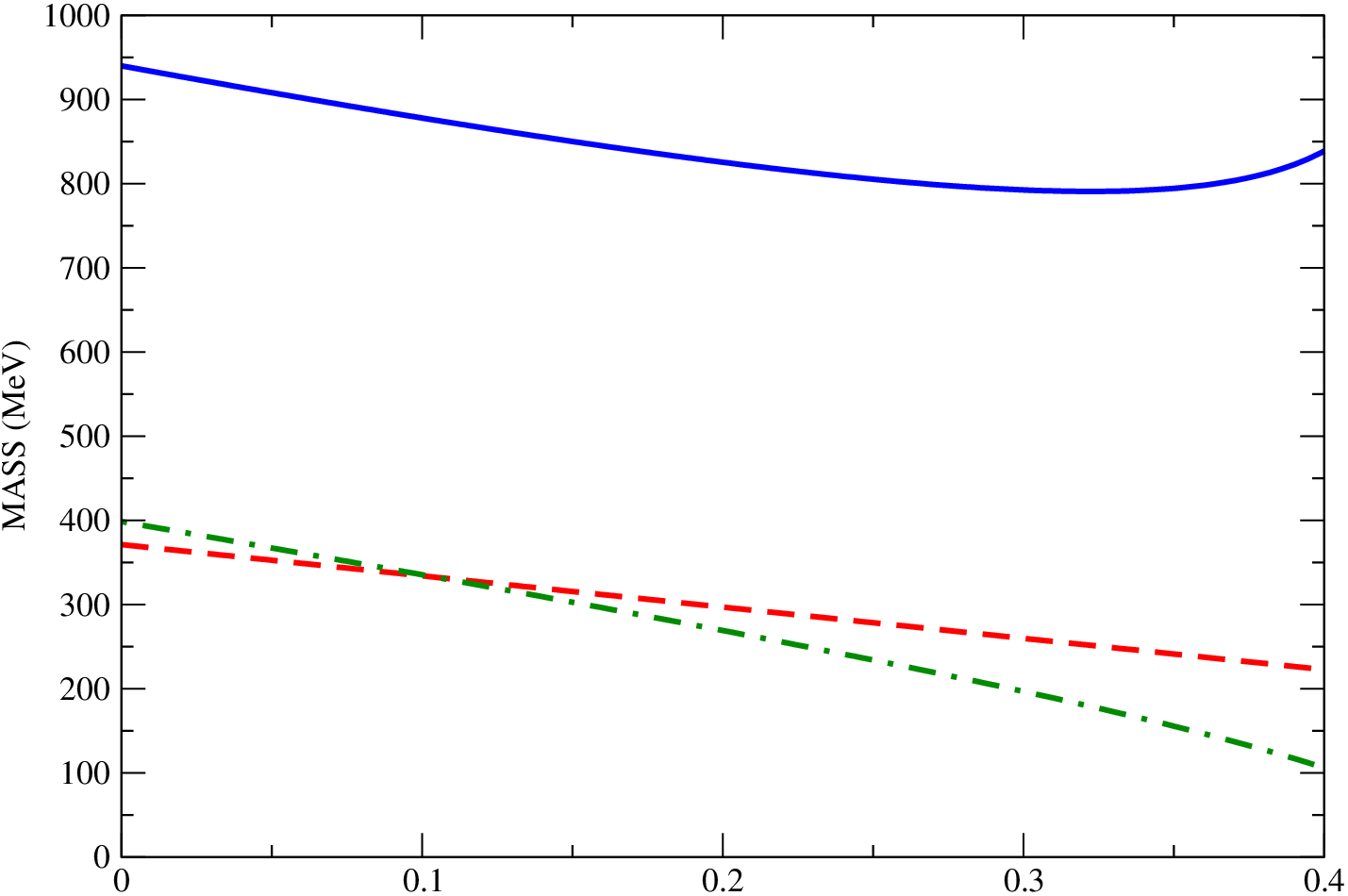}
  \end{minipage}
  &
  \begin{minipage}{.50\linewidth}
	\includegraphics[scale=0.3]{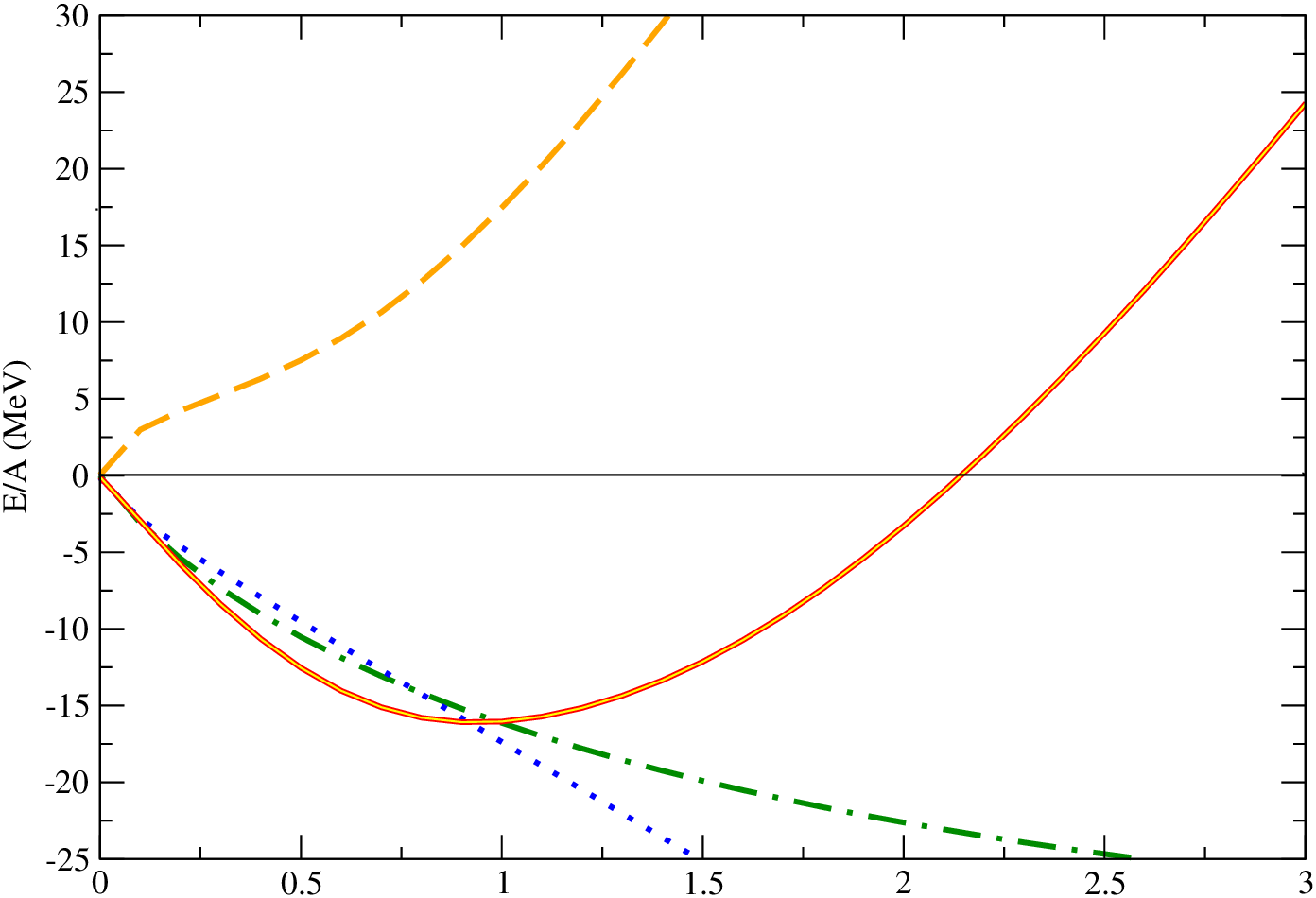}
   \end{minipage}
   \end{tabular}  
\caption{Left panel: Mass of the quark (dashed line), of the diquark (dot-dashed line) and of the nucleon (full line) versus the relative deviation of the scalar field with respect to its vacuum value. Right panel: Binding energy of nuclear matter versus nuclear matter density in units of normal density. 
The full line corresponds to the full result and the dashed line represents the  Hartree result. 
The dot-dashed line corresponds to 
the contribution of the Fock term and the  dotted line  represents the
correlation energy. All the numerical inputs are given in the text.}
\end{figure} 

\subsection{Effect of confinement: simple  models for the in-medium nucleon}
We now come to the last point of this paper, namely  the modelling of the nucleon mass origin and the scalar
response of the nucleon defined from the second derivative of the nucleon mass with respect to the scalar field.
For a nucleon made of the simple adjonction of three NJL constituent quarks (or a NJL quark and a NJL diquark) the scalar coupling constant is independant of the scalar field and there is no  scalar response. The importance of the response is related to the respective roles of chiral symmetry breaking and confinement in the generation of the nucleon mass. In a previous work \cite{EC07} we have introduced a model of a nucleon made of three constituent
quarks bound together by a confining harmonic force. The magnitude of the scalar response which followed was 
too small to prevent the collapse of nuclear matter. We will come back later to this 
type of model. A possibility of improvement is to reduce the relative role of chiral symmetry breaking. This can
be achieved  by considering a nucleon made of a quark and a sufficiently light diquark to leave enough room for
confinement. A practical  advantage is that a three-body problem is transformed into a  simpler two-body problem. Beside this simplification,
there are theoretical and phenomenological reasons to favor a quark-diquark model of the nucleon with relatively
light scalar-isoscalar diquark. For instance the work of Shuryak {\it et al} on hadronic current-correlation 
functions based on a random instanton vacuum \cite{SH92} finds a strong attraction in the scalar-isoscalar 
channel leading to a diquark with a mass about $400\,MeV$.

As discussed in a set of works of Bentz et al (see ref. \cite{BT01}), it is possible to construct   from the NJL model a nucleon  which  has a diquark component.  The diquark mass   is also medium dependent since it depends on the constituent quark mass.  If   we take for the the coupling constant in the diquark channel, 
$\tilde{G}_1=0.92\, G_1$, we obtain for the vacuum diquark  $M_D= 398.5\, MeV$
which turns out to be nearly equal to the constituent quark mass. In \cite{BT01}, a nucleon scalar response is obtained through the inclusion of  an infrared cutoff $\mu_R\simeq\, 200 MeV$  in the Schwinger proper time regularization scheme. Such a prescription implies that  quarks cannot propagate at relative distance larger than $1/\mu_R$, hence mimicking a confinement mechanism. Here we propose to incorporate confinement in a more direct way. Since the diquark is in an anti-triplet color state, it is physically plausible that a string develops between the quark and the diquark as in a $Q\bar Q$ meson. We thus introduce a confining potential between the quark and the diquark: $V(r) = K\, r^2 / 2$.
In the non relativistic limit, the problem reduces to solve the Schrodinger equation for a particle with reduced mass $\mu$, placed in an harmonic potential. In this limit the mass of the (in-medium) nucleon is given by~:
\begin{equation}
M_N(\bar{\cal S})=M(\bar{\cal S})\,+\, M_{D}(\bar{\cal S})\,+\,\frac{3}{2}\sqrt{\frac{K}{\mu(\bar{\cal S})}}\qquad\hbox{with}\qquad
\mu=\frac{M\, M_D}{M + M_D}
\end{equation}
Taking for the string tension a standard value $K=(290 \, MeV)^3$, we obtain for the vacuum nucleon mass $M_N=1304\, MeV$. The nucleon mass origin splits roughly into a chiral symmetry breaking component ($60\%$) and a confinement
component ($40\%$). The vacuum value scalar coupling constant of this nucleon to the effective scalar field is  $g_\sigma \equiv(g_S)_{eff}(M_0)=7.14$. This leads to the value of the non pionic piece of the sigma term calculated in the model~:  $\sigma^{(no pion)}_{N \sigma}=F_\pi g_\sigma\,\left(M^{2}_{\pi}/(M^{2}_{\sigma})_{eff}\right)=
30\, MeV$, as was required. In the linear sigma model where $g_\sigma=M_N/F_\pi=10$  we recover the Ioffe sum rule just corrected from pionic effects. With confinement the value of the scalar coupling constant is reduced with respect to the linear sigma model or additive NJL model  ($g_\sigma=M_N/F_\pi=10$) and the mass evolution is slower than the condensate one (see eq.\, (\ref{COND},\ref{MASS})). The suppression of the pionic contribution to the mass evolution further accentuates the difference between the mass and condensate evolutions. For instance at normal nuclear density the condensate has dropped by $\simeq 30\%$.  With the value $g_\sigma\simeq 7$ deduced above, the mass reduction is significantly lower, $ \simeq 13\%$.

\begin{figure}
\includegraphics[scale=0.3,angle=0]{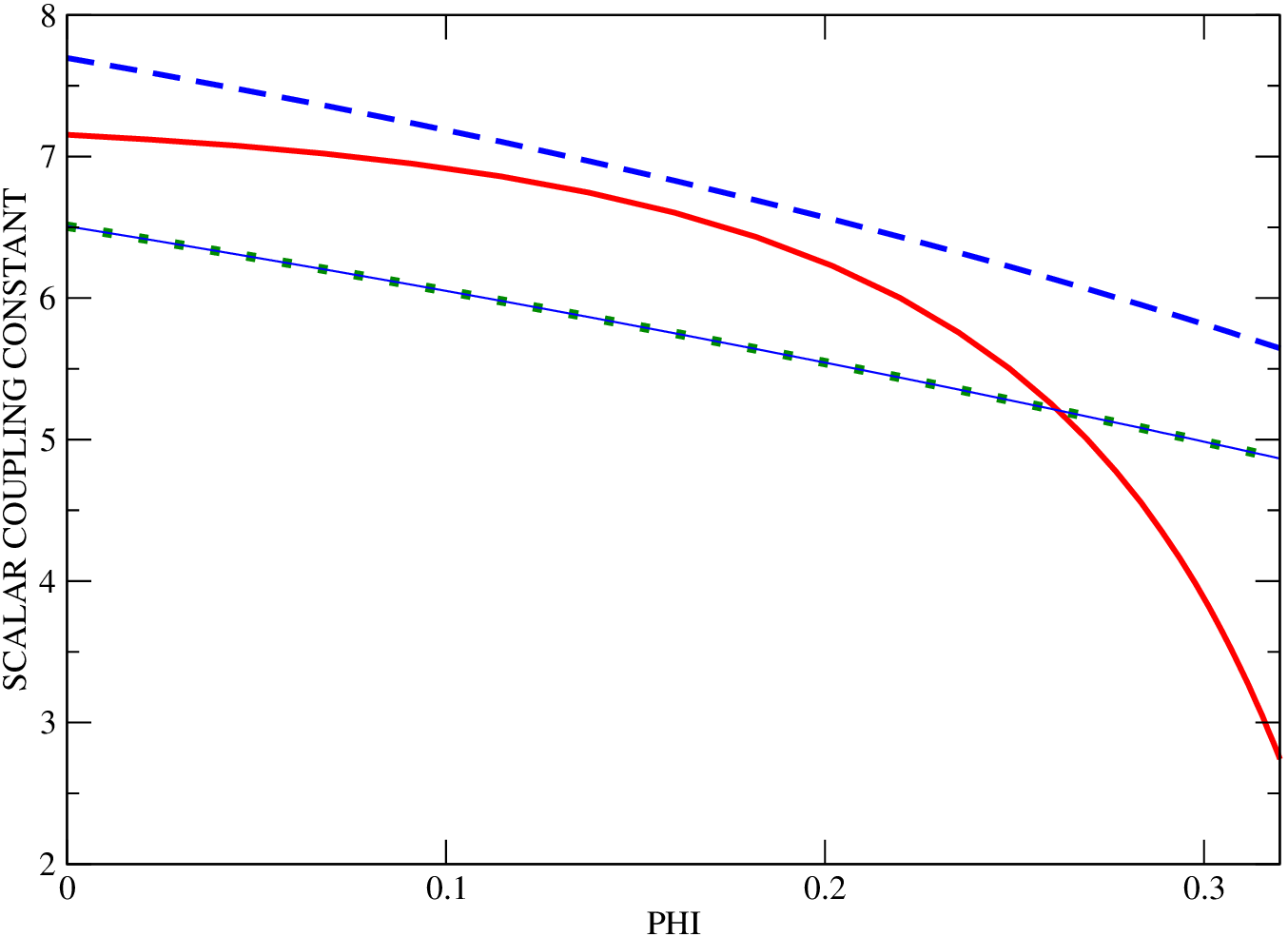}
\label{fig:scalarcoupl}
\caption{effective scalar coupling constant  versus the relative deviation, $\Phi=(M_0-M)/M_0\equiv \left|\bar{s}\right|/F_\pi$, of the scalar field with respect to its vacuum value for the linear confining potential (dashed line), the quadratic linear potential (dotted line) and for the quark-diquark model (full line).}
\end{figure}
In order to show that such a model is capable of describing the saturation properties of nuclear matter we calculate the energy of symmetric nuclear matter in the Hartree approximation, using eq.(\ref{EOS}, \ref{MIN}).
The resulting curve displays a saturation mechanism driven by the scalar nucleon response  ($\kappa_{NS}$, proportional to the second derivative of the nucleon mass with respect to $\bar{\cal S}$) which has a positive value. Said differently the scalar coupling constant, $\partial M_N/\partial\bar{\cal S}$, is a decreasing function of $\left|\bar{s}\right|$ or the density. This translates into the fact that the nucleon mass stabilizes or even increases with increasing $\left|\bar{s}\right|$ (see fig. 1).  However the binding is nevertheless not sufficient unless we decrease artificially the vector coupling constant $G_2$ at a value much smaller than the VDM result. In order to improve the description, although this is not necessarily consistent with our present nucleon model, we add on top of the Hartree mean field result the pion loop (Fock term and correlation energy) contribution obtained in our previous work \cite{CE07}. Taking the value of $G_2$ at the value quoted previously, 
$G_2=0.78\,(G_2)^{VDM}$, we obtain a  decent saturation curve shown in fig. 1. Likely a fully consistent calculation 
within the model of the pion loop energy would  modify the result but a fine tuning on $G_2$ would be presumably 
sufficient to recover the correct saturation curve. The lesson of this simple model calculation seems to confirm our previous conclusions. The confinement effect (scalar response of the nucleon) is able  to stabilize nuclear matter and the pion loop correlation energy helps to get the correct binding energy.

We have shown that an acceptable quark-diquark model of the nucleon makes plausible the role of the background scakar field in the nuclear binding. It is interesting to investigate if other confining mechanisms can achieve the same result. For this we have also studied models where the nucleon is made of three constituent quarks moving  in a mean-field linear confining potential but shifted with a constant attractive potential mimicking  short range attraction~:
$V=\left[(1 + \gamma_0)/{2}\right]\left(K_2\, r\,-\,2\,V_0\right)$. 
This model has been successfully utilized for baryon spectroscopies studies by Jena {\it et al} \cite{JBP97}. We do not aim to justify this particular equally mixed scalar and vector confining potentials, the main motivation being the existence of analytical solutions. 
 The energy of the lowest orbit, solution of the Dirac equation, is~: 
$$ E(M)=M\,-2\,V_0\,+\,\sqrt{K_2}\,x_q\,\quad\hbox{with}\,x_q\,\hbox{solution of}\quad x^{4}_{q}+ 2
\frac{M-V_0}{\sqrt{K_2}}x^{4}_{q}- (2.33811)^3=0$$
and the mass of the in-medium nucleon (in absence of CM correction) is $M_N(\bar{\cal S}=M)=3E(M)$.
Hence the quark mass contribution (essentially the chiral symmetry breaking contribution) to the quark orbital energy  and then to the nucleon is reduced due to the presence of the attractive shift, $-2 V_0$, leaving more room for the confining part. It is possible to show that the scalar coupling constant (still omitting CM correction) can be written as
$(g_S)_{eff}(\bar{\cal S})=3 \,(M_0/F_\pi)\, q_s$
where $q_s=\int d^3 r\,\left(u^2-v^2\right)(r)$ is the quark scalar charge. We see that the scalar field contribution to the sigma term is represented by the usual integrated scalar quark density as in bag models. In practice we also include in the numerical 
calculation the effect of CM correction using the results quoted in ref. \cite{JBP97}. If we  take $K_2 =( 300\, MeV)^2$ and  $V_0=200\,MeV$ it is possible to obtain a saturation curve but the saturation has the tendency to come too early. Certainly this point deserves a more detailed study. Here we wish to concentrate on the main result, namely a decreasing scalar coupling constant when increasing $\left|\bar{s}\right|$ as demonstrated by the dashed curve on fig. 2. We also checked that replacing the linear potential by a quadratic potential,
$V=\left[(1 + \gamma_0)/{2}\right]\left(K_3\, r^2\,-\,2\,V_0\right)$  with  $2K_3 =( 300\, MeV)^3,\, V_0=200\,MeV$,
one obtains similar results as depicted on fig. 2 (dotted curve). It is worthwile to notice that this model differs from the one used in \cite{EC07} by the introduction of the constant attractive shift $-2\, V_0$. 
Again this shift allows to reinforce the role of confinement in the origin of the nucleon mass. 
Also shown on fig. 2 is the behavior of the scalar coupling constant 
for  the quark-diquark model. In this case, the decrease at low density is less strong which translates into a softer equation of state. According to a preliminary study based on a variationnal relativistic calculation the strong dropping beyond  $\left|\bar{s}\right|/F_\pi\approx 0.2$ (which roughly corresponds to normal density) might be to some extent an artefact of the non relativistic approximation.

%


\bibliographystyle{aipproc}   


\IfFileExists{\jobname.bbl}{}
 {\typeout{}
  \typeout{******************************************}
  \typeout{** Please run "bibtex \jobname" to optain}
  \typeout{** the bibliography and then re-run LaTeX}
  \typeout{** twice to fix the references!}
  \typeout{******************************************}
  \typeout{}
 }


\end{document}